\documentclass{article}
\usepackage{hyperref}
\usepackage{bm}
\usepackage{amssymb}	% Extra maths symbols
\usepackage{authblk}
\usepackage{graphicx}
\usepackage{booktabs}
\usepackage{tabularx, booktabs, makecell, caption}
\usepackage{siunitx}
\usepackage{float}\usepackage[
singlelinecheck=false % <-- important
]{caption}

\begin{document}
\title{\bf Influence of plasma on the observational appearance of rotating black holes in Horndeski gravity}
\author{{Malihe Heydari-Fard%
\thanks{Electronic address: \href{mailto:heydarifard@qom.ac.ir}{heydarifard@qom.ac.ir}} and Mohaddese Heydari-Fard\thanks{Electronic address: \href{mailto:mo.heydarifard@qom.ac.ir}{mo.heydarifard@qom.ac.ir}} }\\ {\small \emph{ Department of Physics, The University of Qom, 3716146611, Qom, Iran}}
}

\maketitle

\begin{abstract}
Exploring the influence of plasma on the light rays trajectories in the vicinity of black holes is significat since that astrophysical black holes are generally surrounded by a plasma medium. In this work, we analyze the null geodesics in the space-time of rotating hairy Horndeski black holes immersed in a plasma medium using the Hamilton-Jacobi method. By considering both uniform and non-uniform plasma distributions, we perform a comparative study of their effects on the black hole shadow. The impact of the hair parameter and the inclination angle on the black hole shadow is also investigated. Finally, the shadow results are compared with observational data of the Event Horizon Telescope for M87* to constrain the plasma parameter.
\vspace{5mm}
\end{abstract}

\section{Introduction}

Black holes (BHs) are the most intriguing objects predicted by the theory of General Relativity (GR). As the most powerful gravitational sources, BHs are generally expected to possess substantial angular momentum and magnetic fields. These characteristics establish BHs as ideal laboratories for investigating both matter and gravitational phenomena in astrophysical contexts. Over the past decade, some strong observational evidence has substantiated their existence. For instance, evidence comes from binary star systems that are strong X-ray sources, which indicate the presence of an unseen companion. The mass of this companion, inferred from its gravitational influence on the visible star, is too great for it to be a white dwarf or neutron star \cite{book1}. Furthermore, the detection of gravitational waves from merging binary BHs by the LIGO and Virgo collaborations has definitively confirmed their existence \cite{LIGOScientific:2016aoc}. Another remarkable achievement was the release of the first image of the shadow of M87* \cite{EventHorizonTelescope:2019dse, EventHorizonTelescope:2019ths}, followed by the image of Sgr A* \cite{EventHorizonTelescope:2022wkp}, both captured by the Event Horizon Telescope (EHT) using very long baseline interferometry.

Photons emitted from a luminous source located behind a BH produce what is referred to as the BH shadow - a two-dimensional representation of the dark region from the observer’s celestial perspective. The first studies concerning the shadow produced by static, spherically symmetric BHs were conducted by Synge \cite{Synge:1966okc} and Luminet \cite{Luminet:1979nyg}. They also provided formulas to determine the angular radius and size of the shadow. Bardeen generalized Luminet's paper to address the shadow formed by a rotating Kerr BH \cite{Bardeen}, and showed that the phenomenon of frame-dragging results in a distortion of the shadow's shape. For the study of the BH shadow in alternative gravity theories see \cite{shadow1} and references within it.

Realistic astrophysical BHs are surrounded by a dispersive medium, typically a plasma, through which light rays propagate predominantly. Furthermore, the existence of illuminating sources between the observer and the BH, such as light emitted from an accretion disk, obscure the BH's shadow in the observer’s sky. Thus, analyzing the impact of the surrounding plasma and accretion flow on the trajectory of light rays in such space-times is crucial. One of the tests of GR is the light bending caused by massive compact objects, which results in gravitational lensing and determines the resulting deflection angle and image characteristics. In the past, most of these tests were performed in the weak field approximation of GR; however, recent advances have extended these tests to strong-field regime. The trajectory of a light ray in a vacuum and in a dispersive medium such as plasma near massive compact objects is likely to differ significantly due to refraction and dispersion of the medium \cite{book2}. The influence of a non-magnetize pressureless plasma environment on the shadow of a static spherically symmetric \cite{Perlick:2015vta} and a rotating Kerr BHs was studied in \cite{Perlick:2017fio, 44}. This subject has been extensively studied in different modified gravity theories as well as for various compact objects  \cite{Bisnovatyi-Kogan:2010flt}-\cite{Feleppa:2024vdk}. In recent years, the BHs shadow in the presence of plasma has also been extensively investigated \cite{Abdujabbarov:2015pqp}-\cite{Gohain}. For a review on theoretical models of BHs in the presence of plasma, see \cite{Perlick:2021aok}.

The simplest extension of GR is the scalar-tensor gravity theory, which contains a scalar field $\phi$ alongside the metric tensor $g_{\mu\nu}$ \cite{Damour:1992we}. The most well-known four-dimensional scalar-tensor theory is Horndeski gravity, formulated in 1974 \cite{Horndeski}. This theory characterized by higher derivatives of $\phi$ and $g_{\mu\nu}$ while maintaining second-order field equations, thereby avoiding Ostrogradski instabilities. It also preserves the symmetries of GR, namely diffeomorphism and local Lorentz invariance. In 2011, Deffayet et al. \cite{Deffayet:2011gz} constructed the action for the most general scalar-tensor theories with second-order equations of motion, following the extensions of covariant Kobayashi et al. \cite{Kobayashi:2012wm} subsequently showed that this action coincides with the one originally proposed by Horndeski. In the context of Horndeski gravity theory the hairy BH solutions have been constructed \cite{h1}-\cite{h4} including both the radially dependent \cite{Feng:2015oea}-\cite{Rinaldi:2012vy} and time-dependent scalar field \cite{Minamitsuji:2019shy}-\cite{Babichev:2017lmw}. The rotating hairy Horndeski BHs have also been obtained in reference \cite{h6}. Testing Horndeski gravity with observational data has applied constraints on these theories \cite{h8}-\cite{h16}. In the framework of steady-state Novikov-Thorne model, we have studied the thin accretion disk processes of static and rotating BHs in Horndeski gravity \cite{h18}. The observational appearance of the static hairy Horndeski BHs surrounded by thin disk and thin spherical accretion flows has been studied in \cite{h19}. Rotating Horndeski BH shadows in the absence of plasma has been considered in \cite{h20} and some constraints on the hair parameter $h$ have been obtained. In this paper, we generalize reference \cite{h20} to the realistic case by considering the rotating Horndeski BH in the presence of plasma.

The paper is organized as follows. In Sec.~\ref{2-BH}, we present a brief review of the rotating BHs in Horndeski gravity. In Sec.~\ref{3-shadow}, we introduce the Hamilton-Jacobi equation for photons in a plasma environment, and then study the BH shadow for various plasma distributions. Furthermore, we constrain  the plasma parameters using the observational data from M87* in Sec.~\ref{4-constrain}. Finally, we summarize our results in Sec.~\ref{5-conclusion}.

\section{Rotating BHs in Horndeski gravity}
\label{2-BH}
The action of Horndeski theory reads \cite{Babichev}
\begin{eqnarray}
{\cal S}&=&\int d^4x \sqrt{-g}\left[Q_2(\phi,\chi)+Q_3(\phi,\chi)\Box\phi+Q_4(\phi,\chi)R+Q_{4,\chi}\left((\Box\phi)^2-(\nabla^{\mu}\nabla^{\nu}\phi)(\nabla_{\mu}\nabla_{\nu}\phi)\right)\right]\nonumber\\
&+&Q_5(\phi,\chi)G_{\mu\nu}\nabla^{\mu}\nabla^{\nu}\phi-\frac{1}{6}Q_{5,\chi}\left[(\Box\phi)^3-3(\Box\phi)(\nabla^{\mu}\nabla^{\nu}\phi)(\nabla_{\mu}\nabla_{\nu}\phi)
+2(\nabla_{\mu}\nabla_{\nu}\phi)(\nabla^{\nu}\nabla^{\gamma}\phi)(\nabla_{\gamma}\nabla^{\mu}\phi)\right],
\label{b1}
\end{eqnarray}
where $Q_{i=2,..5}$ are arbitrary functions of the scalar field and kinetic term $\chi=-1/2\partial^{\mu}\phi\partial_{\mu}\phi$, and $G_{\mu\nu}$ is the Einstein tensor. Also, $g$, $R$ and $\phi$ stand for the metric determinant, Ricci scalar and the scalar field, respectively. Here, we consider a subclass of the theory in which $Q_i$ are only functions of the kinetic term, i.e. $Q_i(\chi)$, and $Q_5(\chi)=0$ \cite{Bergliaffa}.

Varying action (\ref{b1}) with respect to $\phi_{,\mu}$ and $g^{\mu\nu}$, the four-current $j^{\mu}$ and field equations can be obtained as follows respectively
\begin{eqnarray}
j^{\mu}&=& -Q_{2,\chi}\phi^{,\mu}-Q_{3,\chi}(\phi^{,\mu}\Box\phi+\chi^{,\mu})-Q_{4,\chi}(\phi^{,\mu}R-2R^{\mu\sigma}\phi_{,\sigma})\nonumber\\
&-&Q_{4,\chi,\chi}[\phi^{,\mu}((\Box\phi)^2-(\nabla_{\alpha}\nabla_{\beta}\phi)(\nabla^{\alpha}\nabla^{\beta}\phi))+2(\chi^{,\mu}\Box\phi-\chi_{,\nu}\nabla^{\nu}\nabla^{\mu}\phi)],
\label{b2}
\end{eqnarray}
and
\begin{equation}
Q_4G_{\mu\nu}=T_{\mu\nu},
\label{b3}
\end{equation}
where
\begin{eqnarray}
T_{\mu\nu}&=&\frac{1}{2}(Q_{2,\chi}\phi_{,\mu}\phi_{,\nu}+Q_2g_{\mu\nu})+\frac{1}{2}Q_{3,\chi}(\phi_{,\mu}\phi_{,\nu}\Box\phi-g_{\mu\nu}\chi_{,\alpha}\phi^{,\alpha}
+\chi_{,\mu}\phi_{,\nu}+\chi_{,\nu}\phi_{,\mu})\nonumber\\
&-&Q_{4,\chi}\left(\frac{1}{2}g_{\mu\nu}[(\Box\phi)^2-(\nabla_{\alpha}\nabla_{\beta}\phi)(\nabla^{\alpha}\nabla^{\beta}\phi)-2R_{\sigma\gamma}\phi^{,\sigma}\phi^{,\gamma}]
-\nabla_{\mu}\nabla_{\nu}\phi\Box\phi\right.\nonumber\\
&+&\left.\nabla_{\gamma}\nabla_{\mu}\phi\nabla^{\gamma}\nabla_{\nu}\phi-\frac{1}{2}\phi_{,\mu}\phi_{,\nu}R+R_{\sigma\mu}\phi^{,\sigma}\phi_{,\nu}
+R_{\sigma\nu}\phi^{,\sigma}\phi_{,\mu}+R_{\sigma\nu\gamma\mu}\phi^{,\sigma}\phi^{,\gamma}\right)\nonumber\\
&-&Q_{4,\chi,\chi}\left(g_{\mu\nu}(\chi_{,\alpha}\phi^{,\alpha}\Box\phi+\chi_{,\alpha}\chi^{,\alpha})+\frac{1}{2}\phi_{,\mu}\phi_{,\nu}(
\nabla_{\alpha}\nabla_{\beta}\phi\nabla^{\alpha}\nabla^{\beta}\phi-(\Box\phi)^2)\right.\nonumber\\
&-&\left.\chi_{,\mu}\chi_{,\nu}-\Box\phi(\chi_{,\mu}\phi_{,\nu}+\chi_{,\nu}\phi_{,\mu})-\chi_{,\gamma}[\phi^{,\gamma}\nabla_{\mu}\nabla_{\nu}\phi
-(\nabla^{\gamma}\nabla_{\mu}\phi)\phi_{,\nu}-(\nabla^{\gamma}\nabla_{\nu}\phi)\phi_{,\mu}]\right)
.\label{b4}
\end{eqnarray}

\begin{figure}[H]
\centering
\includegraphics[width=4.0in]{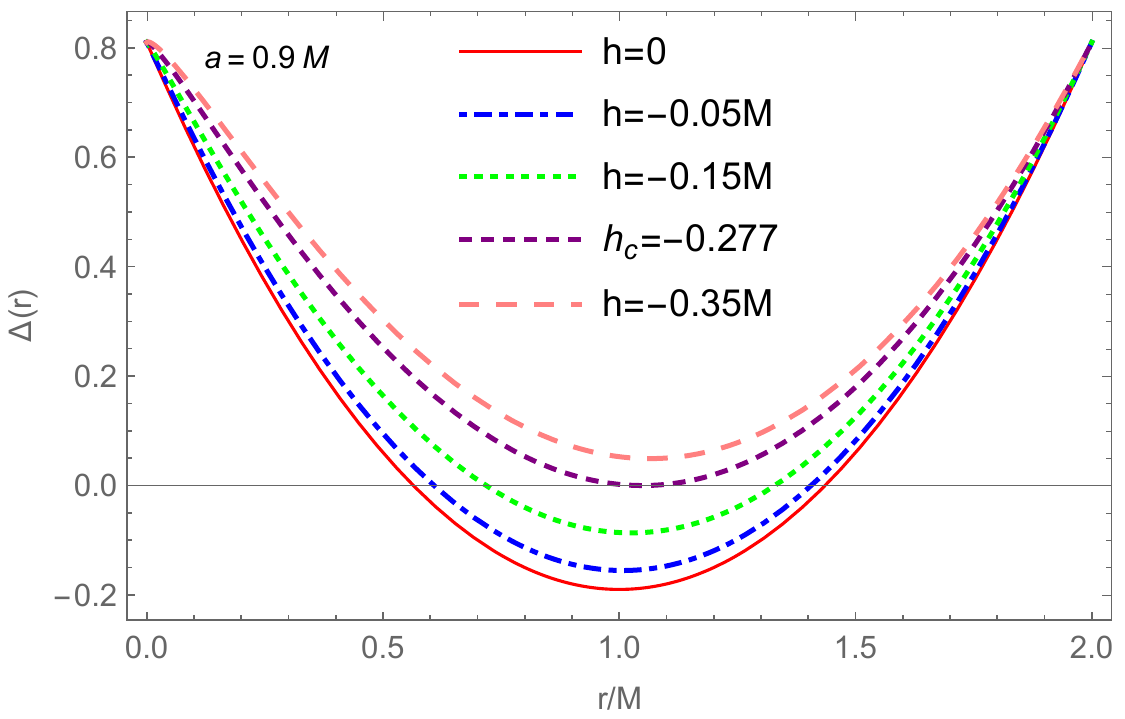}
\caption{\footnotesize The behavior of horizons of the rotating Horndeski BH for $a=0.9M$ with different values of $h$ parameter. The solid curve corresponds to the Kerr BH.}
\label{delta}
\end{figure}

By taking the canonical action of the scalar field $\phi=\phi(r)$, as well as imposing some conditions such as: finite energy of scalar field, i.e., $E=\int_{V}\sqrt{-g}T^0_0d^3x$ and vanishing radial four-current at infinity $j^r=0$, the static and spherically symmetric BH solution of Horndeski gravity has been obtained in \cite{Bergliaffa}
\begin{equation}
ds^2=-f(r)dt^2+\frac{dr^2}{f(r)}+r^2(d\theta^2+\sin^2\theta d\varphi^2),
\label{b5}
\end{equation}
with
\begin{equation}
f(r) = 1-\frac{2M}{r}+\frac{h}{r}\ln\left(\frac{r}{2M}\right),
\label{b6}
\end{equation}
where $M$ and $h$ are the BH mass and the hair parameter, with the dimension of length, respectively. From the condition $f(r)=0$, one can easily find that for $h/M>0$ there is only one horizon at $r=2M$, while in the case of $-2<h/M<0$, there are two horizons; the outer event horizon at $r_+=2M$ and the inner Cauchy horizon at $r_{-}$  \cite{Bergliaffa}; so, we restrict our analysis to this case. Moreover, in the limit $h\rightarrow0$, the above metric becomes the Schwarzschild solution in GR, and it is asymptotically flat. The metric for rotating hairy Horndeski BHs has recently been constructed by applying the revised Newman–Janis algorithm to the static solution (\ref{b5}), which in the Boyer–Lindquist coordinates is given by \cite{Walia}
\begin{eqnarray}
ds^2&=&-\left(1-\frac{2m(r)r}{\Sigma}\right)dt^2-\frac{4am(r)r\sin^2\theta}{\Sigma}dt d\varphi+\frac{\Sigma}{\Delta}{dr^2}\nonumber\\
&+&\Sigma d\theta^2+\sin^2\theta\left[(r^2+a^2)+\frac{2m(r)r a^2\sin^2\theta}{\Sigma}\right]d\varphi^2,
\label{b7}
\end{eqnarray}
with
\begin{equation}
\Delta=r^2+a^2-2m(r)r, \quad \Sigma=r^2+a^2\cos^2\theta.
\label{b8}
\end{equation}
where
\begin{equation}
m(r)=M-\frac{h}{2}\ln\left(\frac{r}{2M}\right).\label{b9}
\end{equation}
The rotating metric (\ref{b7}) is characterized by three parameters $M$, $a$ and $h$, which determines deviation from the Kerr BH. Clearly, in the limiting case $h\rightarrow0$, it reduces to the Kerr metric.
A simple root analysis of $\Delta(r)=0$ shows (see Fig.~\ref{delta}), for a given $a$ there is a critical value of the hair parameter, $h=h_{\rm c}$, such that for $h>h_{\rm c}$ there exist two distinct roots, corresponding to the inner Cauchy horizon, $r_{-}$, and the outer event horizon, $r_{+}$. For the critical value of $h=h_{\rm c}$, two horizons coincide $r_-=r_+$, and we have an extremal BH, while for $h<h_{\rm c}$ there is a naked singularity. Therefore, in what follows we choose the values of $a$ and $h$ parameters so that we have a BH solution.

\section{BH shadow in plasma environment}
\label{3-shadow}
\subsection{Hamilton-Jacobi equation for photons in plasma}
In this section, we aim to explore the shadow of rotating hairy Horndeski BHs within a pressureless and non-magnetized plasma environment. The Hamiltonian for light rays traveling in a given space-time reads \cite{book2}
\begin{equation}
H(x, p) = \frac{1}{2} \left( g^{\mu\nu}(x) p_\mu p_\nu + \omega^2_{p}(x) \right),\label{p1}
\end{equation}
where $\omega_p$ is the plasma electron frequency given by
\begin{equation}
\omega_p^2(x) \equiv \frac{4 \pi e^2}{m_e} N(x),\label{p2}
\end{equation}
with $e$ and $m_e$ are the charge and mass of the electron, respectively while $N(x)$ is the electron density. The plasma frequency $\omega_p$ is related to the refractive index $n$ as follows \cite{Synge}
\begin{equation}
n^2 \equiv 1 - \frac{\omega_p^2(x)}{\omega^2(x)}, \label{p3}
\end{equation}
where $\omega$ is the photon frequency measured by a static observer.
We also assume that the plasma frequency is stationary and axisymmetric, and thus it's not dependent on the coordinates $t$ and $\varphi$. By considering the gravitational redshift effect for the constant of motion $p_t=-\omega_{0}$, the observed redshift frequency is given by $\omega(x)=\frac{\omega_0}{\sqrt{-g_{tt}}}$. Since the refractive index is positive, the following condition should be imposed to the light propagation is possible through the plasma medium \cite{Perlick:2017fio}
\begin{equation}
\omega_0^2 > -g_{tt} \omega_p^2(x). \label{p4}
\end{equation}

Next, we focus on the Hamilton-Jacobi equation for photons as
\begin{equation}
H \left( x, \frac{\partial S}{\partial x^\mu} \right) = \frac{1}{2} g^{\mu\nu}(x) \frac{\partial S}{\partial x^\mu} \frac{\partial S}{\partial x^\nu} + \frac{1}{2} \omega_p^2(x) = 0,\label{p5}
\end{equation}
and consider the separation ansatz
\begin{equation}
S = -\omega_0 t + L\varphi + S_r(r) + S_\theta(\theta), \label{p6}
\end{equation}
where $L$ is the conserved quantity corresponds to the angular momentum of the photon. By substituting (\ref{p6}) in the Hamilton-Jacobi equation, we obtain
\begin{equation}
\Delta \left( \frac{dS_r}{dr} \right)^2 - \frac{1}{\Delta} \left[ (r^2 + a^2) \omega_0 - aL \right]^2 + (L - a\omega_0)^2 + \left( \frac{dS_\theta}{d\theta} \right)^2 - a^2 \omega_0^2 \cos^2 \theta + L^2 \cot^2 \theta + \Sigma \omega_p^2 = 0, \label{p7}
\end{equation}
where $\Delta$ and $\Sigma$ are defined by Eq.~(\ref{b8}). The above equation is separable if and only if we consider the following general form of the plasma frequency as \cite{Perlick:2017fio}
\begin{equation}
\omega_p^2 = \frac{f_r(r) + f_\theta(\theta)}{\Sigma}, \label{p8}
\end{equation}
where $f_r(r)$ and $f_\theta(\theta)$ are arbitrary functions of the coordinates $r$ and $\theta$, respectively. This generalized form of the plasma frequency allows us to investigate the effect of plasma on the dynamics of photons by considering its radial and latitudinal variations. Using Eq.~(\ref{p7}) and Eq.~(\ref{p8}), we find the following expressions
\begin{eqnarray}
\Delta^2\left(\frac{dS_r}{dr}\right)^2 &=& \left[ (r^2 + a^2)\omega_0 - aL \right]^2 - \Delta \left[ {\cal K} + (L - a\omega_0)^2 + f_r(r) \right]\\
\left(\frac{dS_\theta}{d\theta}\right)^2 &=& {\cal K} - L^2 \cot^2\theta + a^2 \omega_0^2 \cos^2\theta - f_\theta(\theta), \label{p9}
\end{eqnarray}
where ${\cal K}$ is the Carter constant. Now, from the Hamilton equations, with $\dot{x^{\mu}}=p^{\mu}=g^{\mu\nu}p_{\nu}$, where the dot denotes derivative with respect to the affine parameter, and $p_{\nu}=\frac{\partial S}{\partial x^{\nu}}$, we can obtain the equations of motion

\begin{eqnarray}
\Sigma \dot{t} &=& \frac{\omega_0}{\Delta} \left[ (r^2 + a^2)^2 -\Delta a^2 \sin^2 \theta- 2 m(r) r a \xi \right],\label{p10}\\
\Sigma \dot{\varphi} &=& \frac{\omega_0}{\Delta} \left[ 2 m(r) r a + \frac{\Sigma - 2m(r)r }{\sin^2 \theta}\xi \right],\label{p11}\\
\Sigma \dot{r} &=& \pm\sqrt{ R(r)},\label{p12}\\
\Sigma\dot{\theta}&=&\pm\sqrt{ \Theta(\theta)},\label{p13}
\end{eqnarray}
with
\begin{eqnarray}
R(r) &=& \omega_0^2\left[ (r^2 + a^2)^2-4m(r)ra\xi + a^2\xi^2 - \Delta\left(\eta+a^2+\xi^2+\frac{f_r(r)}{\omega_0^2}\right)\right],\\
\Theta(\theta) &=& \omega_0^2\left[\eta-\xi^2 \cot^2\theta +a^2\cos^2\theta-\frac{f_\theta(\theta)}{\omega_0^2}\right],\label{p14}
\end{eqnarray}
where we defined two impact parameters as $\eta=\frac{{\cal K}}{\omega_0^2}$ and $\xi=\frac{L}{\omega_0}$. The parameter $\eta$ denotes the distance from the equatorial plane, while $\xi$ shows the distance from the axis of rotation.

\subsection{Spherical photon orbits}
To study the apparent shape of the Horndeski BH surrounded by plasma, we have to consider the unstable photon orbits satisfying the following conditions
\begin{equation}
R''(r)>0, \quad \Theta(\theta)\geq 0,\label{p15}
\end{equation}
where prime refers to the derivative with respect to the radial coordinate. We initially aim to identify the critical photon orbits corresponding to the boundary of the shadow. To this end, we use the impact parametrs $\eta$ and $\xi$, which have a crucial role in determining the boundary of the shadow. It should be noted that these critical orbits correspond to the most unstable circular trajectories, which are associated with the maximum of the effective potential $V_{\rm eff}$. Therefore, the conditions of  the existence of these unstable circular orbits are given by
\begin{equation}
V_{\rm eff}(r)|_{r=r_{\rm c}} = 0, \quad V'_{\rm eff}(r)|_{r=r_{\rm c}} = 0,\label{p16}
\end{equation}
where $r_{\rm c}$ represents the critical photon orbit. Since the radial geodesic equation (\ref{p12}) can be expressed as $\dot{r}^2 + V_{\rm eff} = 0$, the above conditions can be rewritten as
\begin{equation}
R(r)|_{r=r_{\rm c}} = 0, \quad R'(r)|_{r=r_{\rm c}} = 0.\label{p17}
\end{equation}
By solving these conditions, one can find the following expressions for impact parameters
\begin{eqnarray}
\eta &=&\left[ \frac{1}{\Delta} \left((r^2 + a^2)^2 - 4 a m(r) r \xi + a^2 \xi^2 \right) - \left(a^2 + \xi^2 + \frac{f_r}{\omega_0^2} \right)\right]_{r=r_c},\label{p18}\\
\xi &=& \frac{-B - \sqrt{B^2 - 4AC}}{2A},\label{p19}
\end{eqnarray}
where
\begin{eqnarray}
A&=&\frac{\Delta'}{\Delta} a^2,\\
B&=&4am(r)(1-\frac{\Delta'}{\Delta}r)+4am'(r)r\\
C&=& -4r(r^2 + a^2) + \frac{\Delta'}{\Delta}(r^2 + a^2)^2 + \Delta \frac{f_r'}{\omega_0^2}.\label{p20}
\end{eqnarray}
One can easily check by setting $f_r(r)=0$ and turning off Horndeski parameter $h$ in $\Delta(r)$ function, expressions (\ref{p18}) and (\ref{p19}) reduce to the following expressions for the Kerr BH in the vacuum case
\begin{eqnarray}
\xi &=&\frac{r^2(r-3M)+a^2(r+M)}{a(M-r)},\label{p21}\\
\eta &=& \frac{r^3(4Ma^2-r(r-3M)^2)}{a^2(M-r)^2}.\label{p22}
\end{eqnarray}

Note although the boundary of the BH shadow can be completely determined by $\xi$ and $\eta$, but to observe the shadow in the observer's sky, we have to find the celestial coordinates
\begin{eqnarray}
\alpha &=& \lim_{r_{\rm o} \to \infty} \left( -r_{\rm o}^2 \sin \theta_{\rm o} \frac{d\varphi}{dr} \right),\label{p23}\\
\beta &=& \lim_{r_{\rm o} \to \infty} \left(r^2 \frac{d\theta}{dr}\right),\label{p24}
\end{eqnarray}
where $r_{\rm o}$ is the observer distance to the BH and $\theta_{\rm o}$ is the inclination angle. Using geodesic equations (\ref{p10})-(\ref{p13}), we can obtain the expression for the celestial coordinates and thus investigate the plasma effects on the apparent shape of the Horndeski BHs. In what follows, we consider both  the homogeneous and inhomogeneous plasma distributions.

\subsection{Shadow of Horndeski BHs in plasma space-time}
Now, we are ready to study the plasma profiles in the vicinity of rotating hairy Horndeski BHs and investigate its effects on the BH shadow. An interesting case was introduced by Shapiro et al. \cite{Shapiro} showing that the squared plasma frequency is proportional to $r^{-3/2}$ and independent of $\theta$ for non-magnetized pressureless plasma. However, for this plasma distribution the separability condition (\ref{p8}) is not satisfied; Therefore, we will take an additional $\theta$ dependence of the plasma frequency to satisfy the separability condition \cite{Perlick:2017fio}.

$\bullet $ {\bf Case A: Homogeneous plasma distribution}

\begin{figure}[H]
\centering
\includegraphics[width=2.20in]{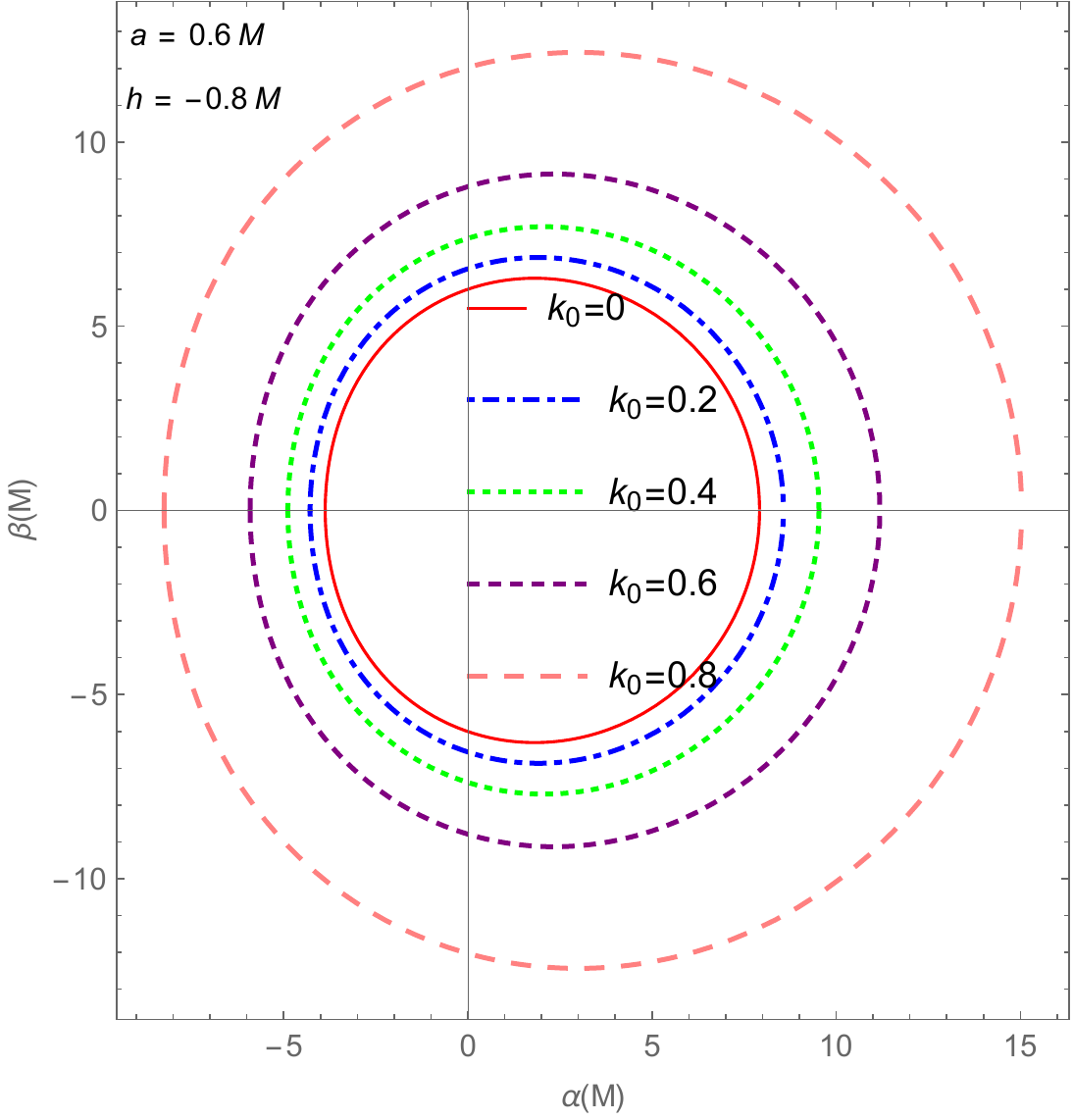}
\includegraphics[width=2.20in]{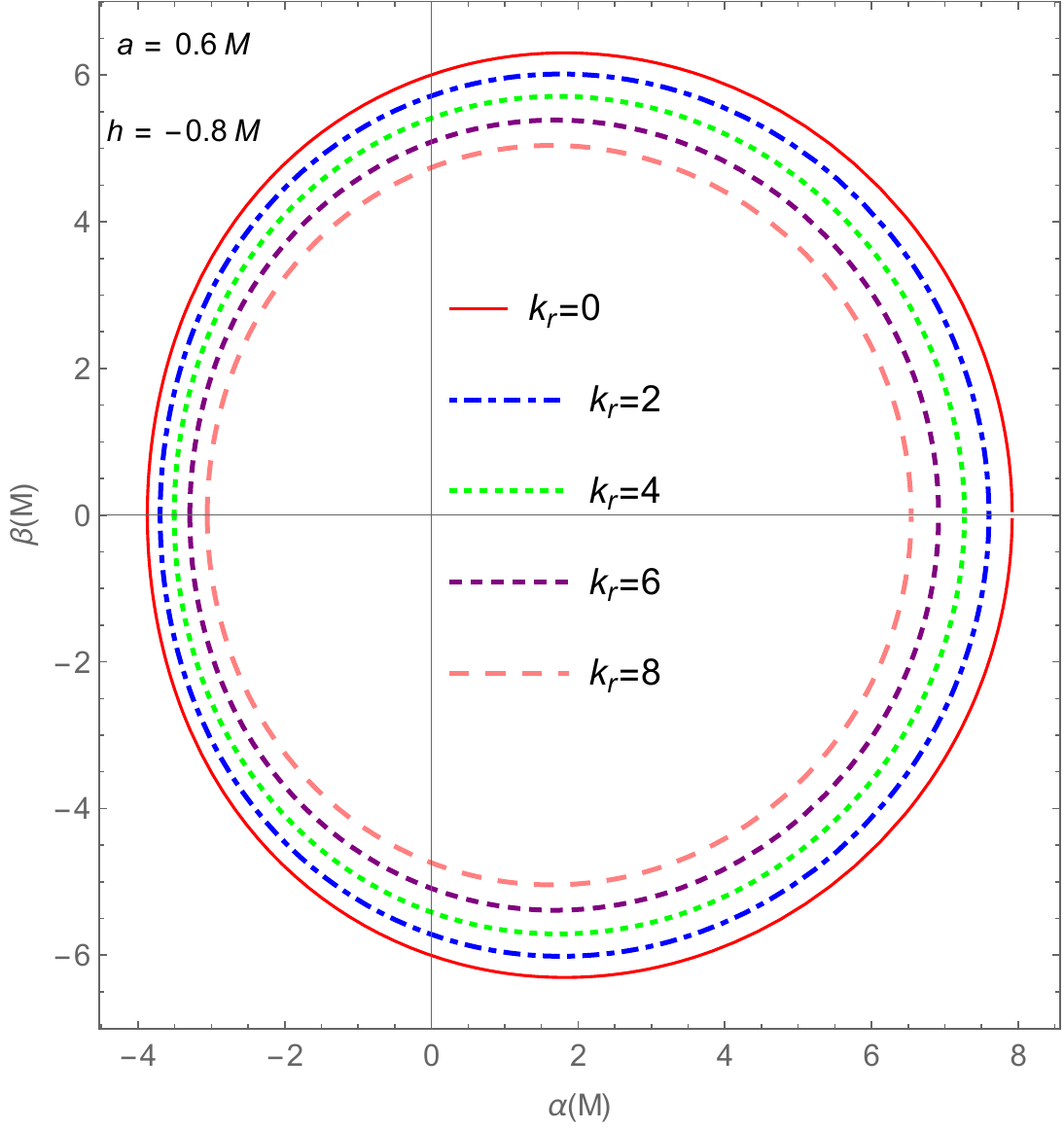}
\includegraphics[width=2.20in]{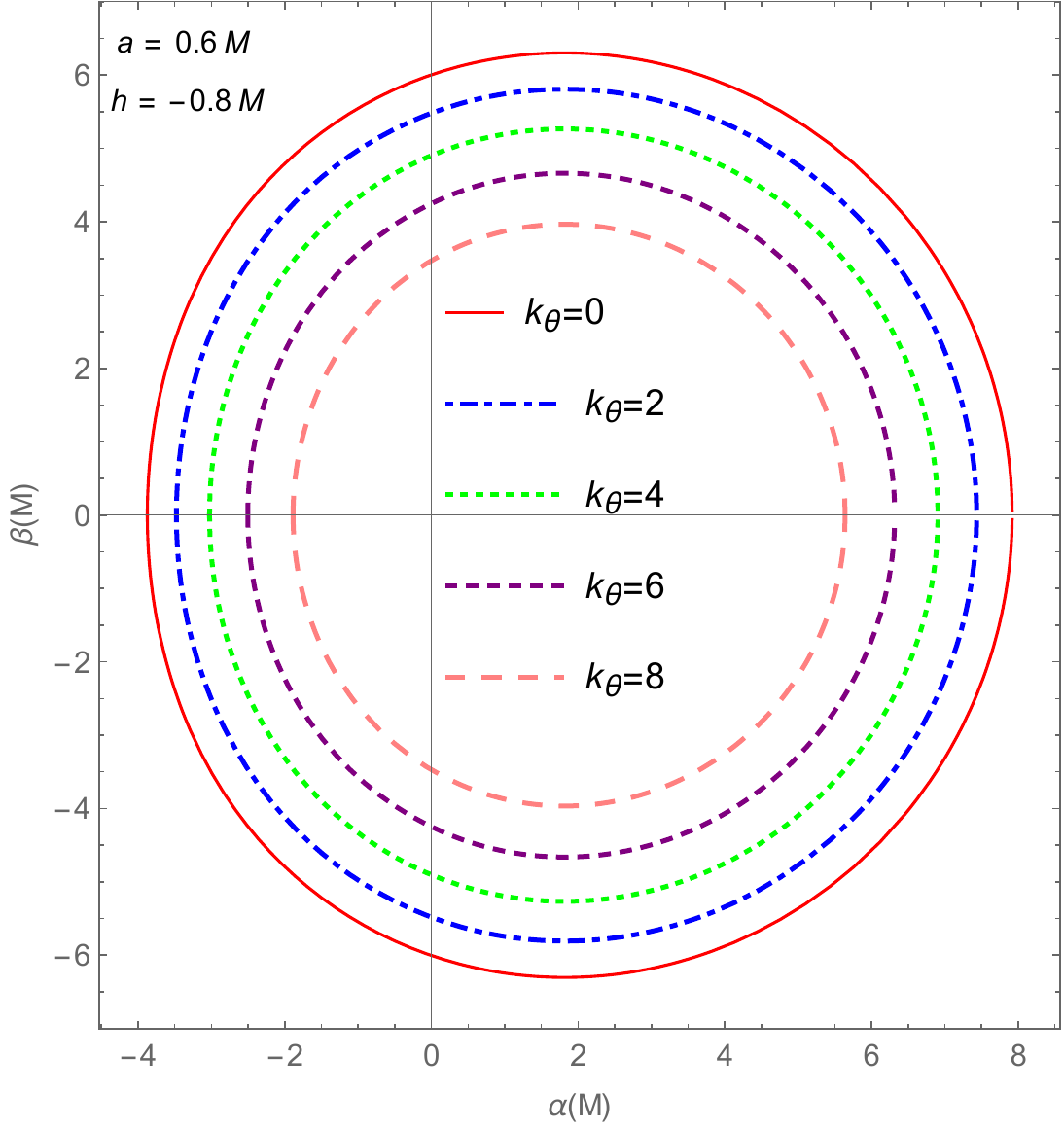}\\
\includegraphics[width=2.20in]{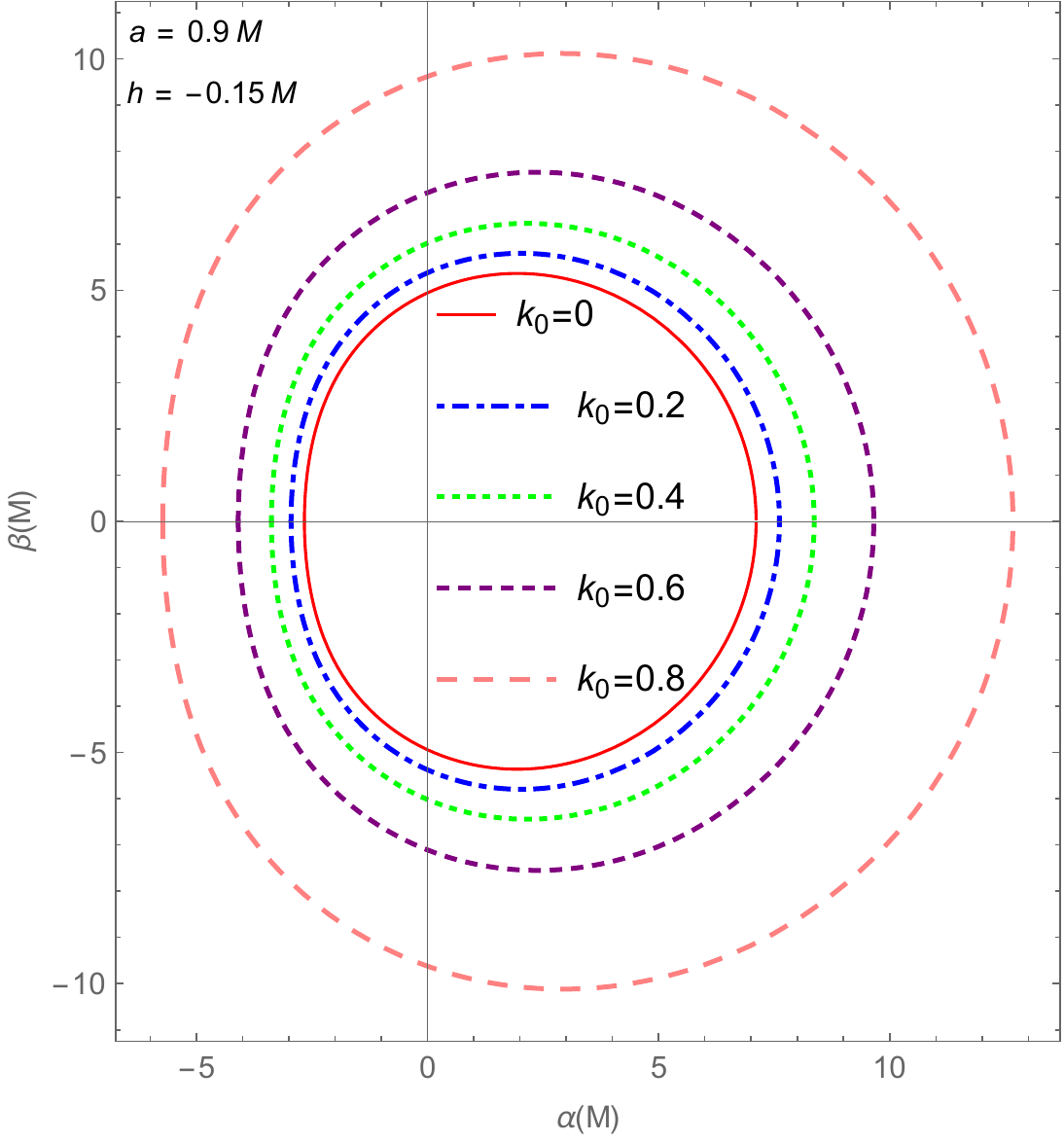}
\includegraphics[width=2.20in]{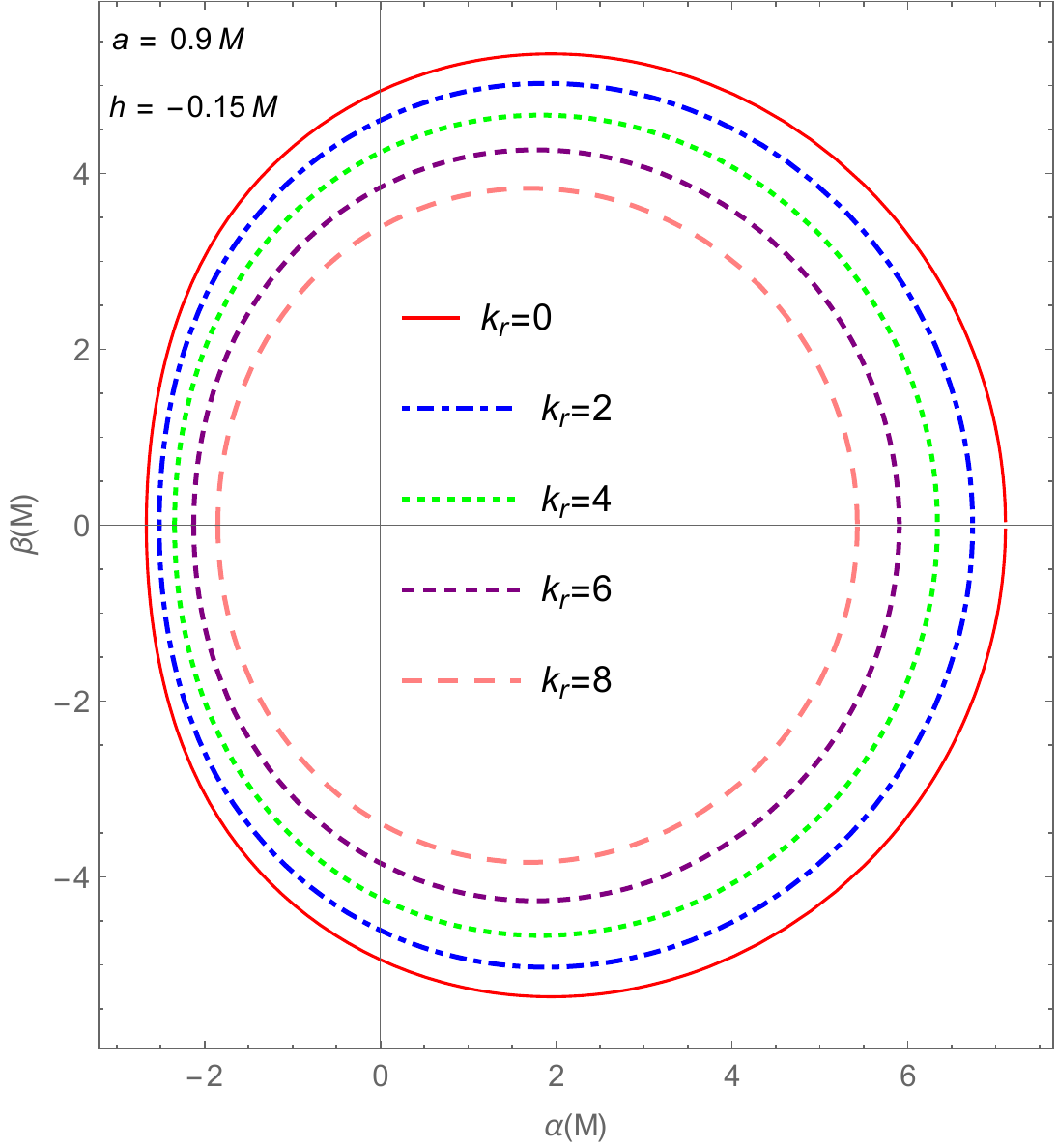}
\includegraphics[width=2.20in]{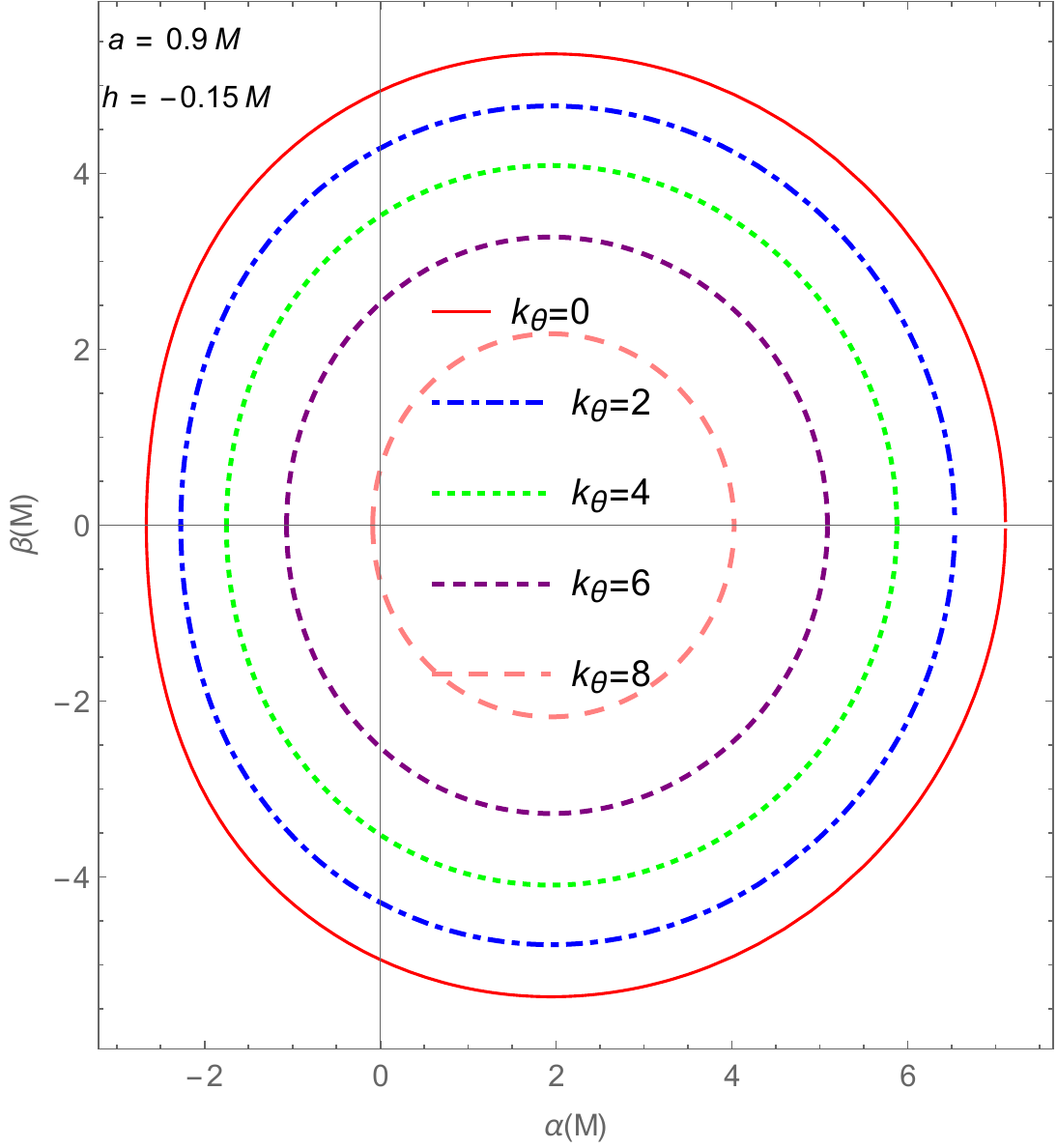}
\caption{\footnotesize Horndeski BH shadows with various plasma parameters $k_0$, $k_r$ and $k_{\theta}$: homogeneous plasma distribution with $\omega_p^2= k_0\omega_0^2 $ (left panel), inhomogeneous distribution with $\omega_p^2 = \frac{k_r\sqrt{r}}{r^2+a^2\cos^2\theta}\omega_0^2$ (middle panel) and with $\omega_p^2 = \frac{k_{\theta}(1+2\sin^2\theta)}{r^2+a^2\cos^2\theta}\omega_0^2$ (right panel). In each panel the solid curve corresponds to the vacuum case. The inclination angle is set to $\theta_{\rm o}=\frac{\pi}{2}$.}
\label{shadow}
\end{figure}

First, we will consider an ideal situation which is homogeneous plasma distribution and is generally given by:
\begin{equation}
\frac{\omega_p^2}{\omega_0^2} = k_0,\label{p25}
\end{equation}
where $k_0$ is the homogeneous plasma parameter, and to hold condition (\ref{p4}) it needs to be in the interval $(0,1)$. From Eqs. (\ref{p8}) and (\ref{p25}) we find
\begin{equation}
f_r = k_0\omega_0^2 r^2 , \quad f_\theta = k_0 \omega_0^2a^2 \cos^2 \theta .\label{p26}
\end{equation}
Then, the celestial coordinates for the case of homogeneous plasma are obtained from Eqs. (\ref{p23}) and (\ref{p24}) along with geodesic equations as follows
\begin{eqnarray}
\alpha &=& -\frac{\xi}{\sin \theta \sqrt{1-k_0}},\nonumber\\
\beta &=& \frac{\sqrt{\eta - \xi^2 \cot^2 \theta + a^2 \cos^2 \theta (1-k_0)}}{\sqrt{1-k_0}}.\label{p27}
\end{eqnarray}
In the left panels of Fig.~\ref{shadow}, we see the effect of homogeneous plasma on the size of the BH shadow, showing that the larger plasma parameter $k_0$, the larger BH shadow will be. So, the Horndeski BH in vacuum with $k_0=0$, has the smallest shadow size. Also, the comparison of left panels shows the D-shaped shadow for high rotation BHs.

$\bullet $ {\bf Case B: plasma distribution with $f_{\theta}=0$}

For the second case, we consider an inhomogeneous plasma profile that the density go as $r^{-3/2}$, as for a dust \cite{Perlick:2017fio}; however the plasma frequency have to expressed as follows to satisfy the separability condition
\begin{equation}
\frac{\omega_p^2}{\omega_0^2} = \frac{k_r\sqrt{r}}{r^2+a^2\cos^2\theta},\label{p28}
\end{equation}
where $k_r$ is the radial plasma parameter, and along with Eq. (\ref{p8}) gives the expression for $f_{r}$ and $f_{\theta}$ as
\begin{equation}
f_r = k_r\omega_0^2 \sqrt{r} , \quad f_\theta = 0.\label{p29}
\end{equation}
Then, using Eqs. (\ref{p23}) and (\ref{p24}), the celestial coordinates for this profile are given by
\begin{eqnarray}
\alpha &=& -\frac{\xi}{\sin \theta},\nonumber\\
\beta &=& \sqrt{\eta - \xi^2 \cot^2 \theta + a^2 \cos^2 \theta}.\label{p30}
\end{eqnarray}
The effect of such plasma profile on the BH shadow, is shown in the middle panels of Fig.~\ref{shadow} for $a=0.6$ and $a=0.9$ with inclination angle $\theta=90^{\circ}$. As is clear, in contrast to the case of homogeneous distribution, the shadow size decrease with increasing the radial plasma parameter. Therefore, there is  an upper limit for the plasma density beyond which the BH shadow vanishes completely.
Moreover, we note that the values for the radial plasma parameter $k_r$ are chosen in such a way that condition (\ref{p4}), as well as conditions (\ref{p15}) and (\ref{p17}) for unstable circular orbits are satisfied.

$\bullet $ {\bf Case C: plasma distribution with $f_{r}=0$}

Finally, we consider a plasma distribution with $f_{r}=0$.  The Hamiltonian separability condition (\ref{p8}) prevents us from choosing a plasma profile that depends entirely on $\theta$. So, we take the following plasma profile as proposed by Perlick et al \cite{Perlick:2017fio}
\begin{equation}
\frac{\omega_p^2}{\omega_0^2} = \frac{k_{\theta}(1+2\sin^2\theta)}{r^2+a^2\cos^2\theta},\label{p31}
\end{equation}
where $k_{\theta}$ is the latitudinal plasma parameter, and from Eq. (\ref{p8}) we find
\begin{equation}
f_r = 0 , \quad f_\theta =k_{\theta}\omega_0^2 (1+2\sin^2\theta).\label{p32}
\end{equation}
The celestial coordinates for this case can be obtained from (\ref{p23}) and (\ref{p24}) as
\begin{eqnarray}
\alpha &=& -\frac{\xi}{\sin \theta},\nonumber\\
\beta &=& \sqrt{\eta - \xi^2 \cot^2 \theta + a^2 \cos^2 \theta - k_{\theta} (1 + 2 \sin^2 \theta)}.\label{p33}
\end{eqnarray}
The shadow of rotating Horndeski BHs for this plasma profile has been displayed in right panels of Fig.~\ref{shadow}, in the case of $a=0.6M$ and $a=0.9M$. As is clear, the effect of this plasma profile on the BH shadow is similar to the previous case, namely by increasing the $k_{\theta}$ parameter, the shadow radius decreases. However, it is easy to see that for the same values of $k_{r}$ and $k_{\theta}$, the shadow radius is smaller for this plasma distribution in comparison to the previous case. Again, it should be noted that the values of $k_{\theta}$ are determined in such a manner that satisfy conditions (\ref{p4}) and (\ref{p15}).

\begin{figure}[H]
\centering
\includegraphics[width=2.00in]{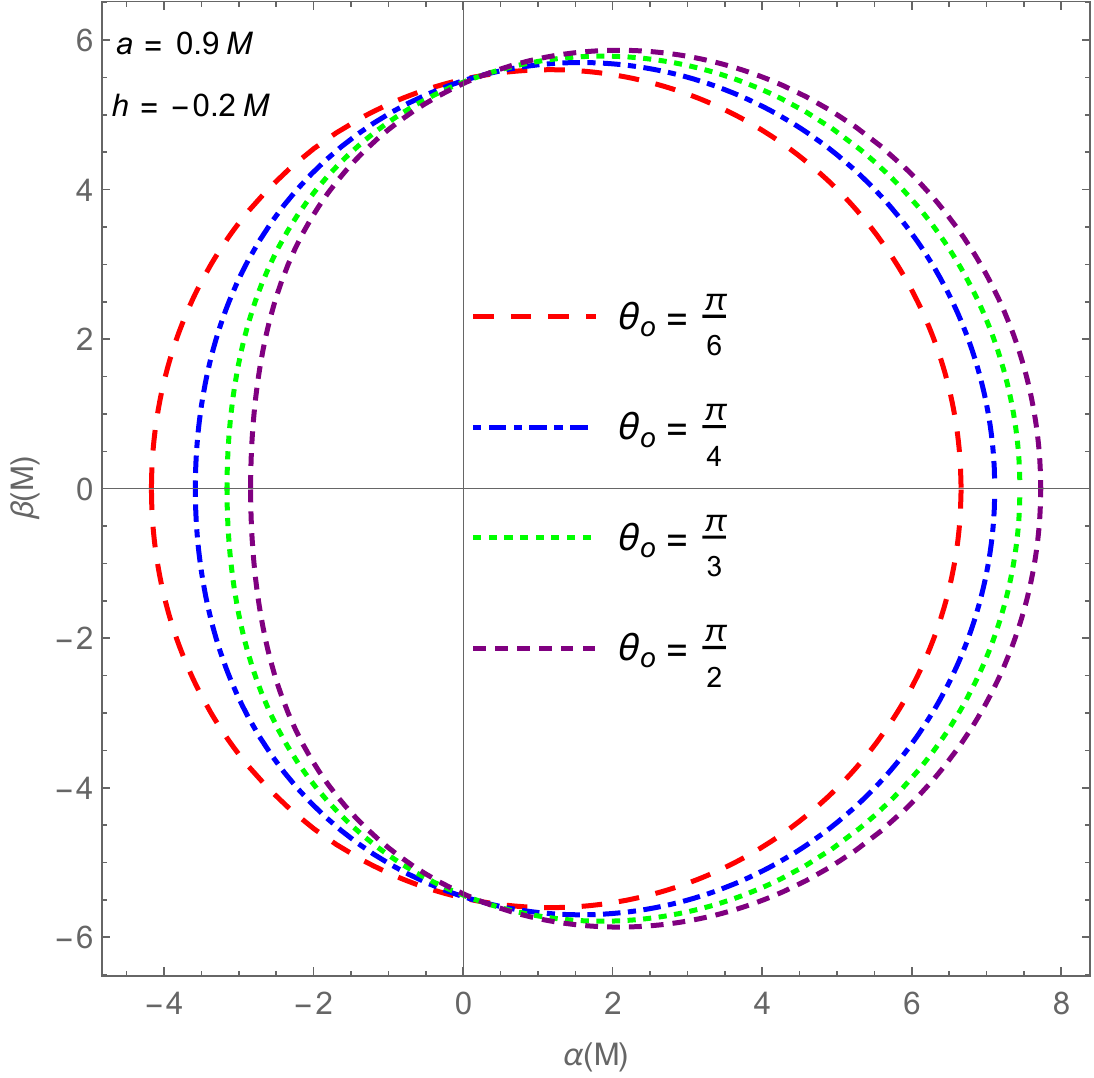}
\includegraphics[width=2.25in]{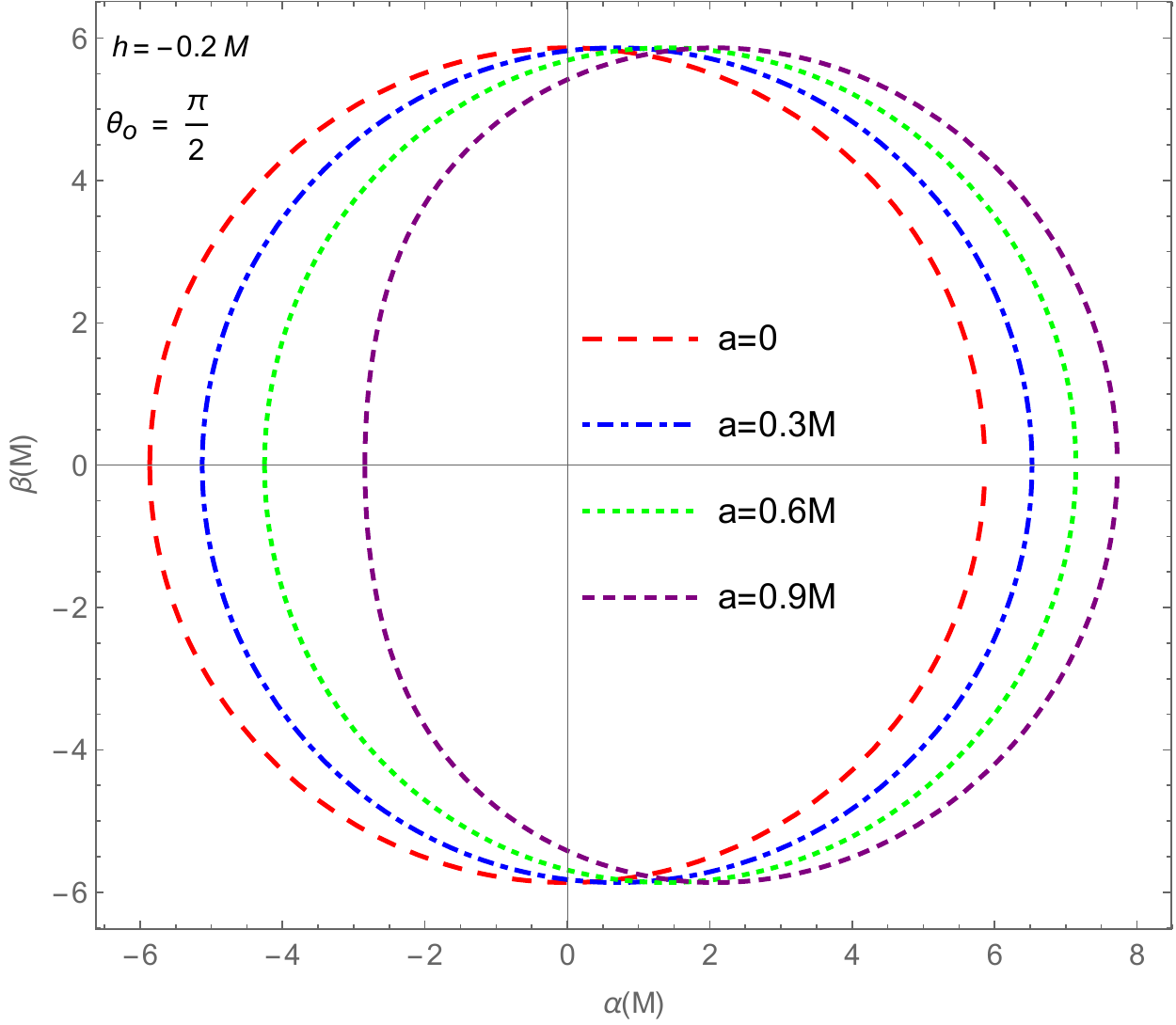}
\includegraphics[width=1.90in]{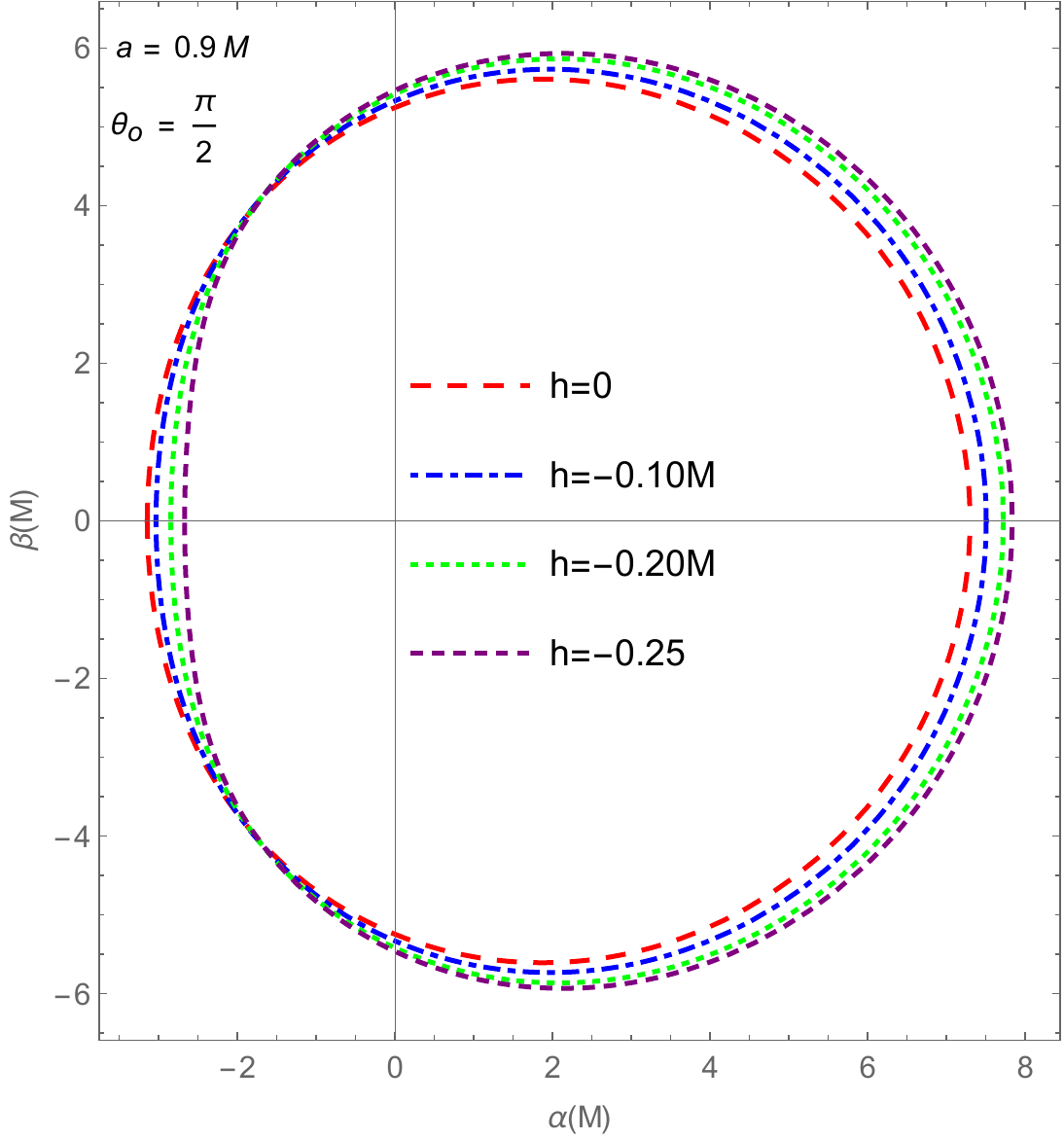}
\caption{\footnotesize Horndeski BH shadows in homogeneous plasma distribution for different inclinations angles with $a=0.9M$ and $h=-0.2M$ (left panel), different spin parameter with $h=-0.2M$ and $\theta_{\rm o}=\frac{\pi}{2}$ (middle panel), and for different Horndeski parameter with $a=0.9M$ and $\theta_{\rm o}=\frac{\pi}{2}$ (right panel). The homogeneous plasma parameter is set to $k_0=0.2$.}
\label{shadow2}
\end{figure}

Considering the homogeneous plasma profile with $k_{r}=0.2$, the shadows of Horndeski BH for different inclination angles, different  values of spin parameters as well as different values of Horndeski parameters have been displayed in Fig.~\ref{shadow2}. We see that by increasing the inclination angle and the spin parameter, the shadow shape deviates from circularity, which the effect is more pronounced for spin parameter. Moreover, from the right panel of the figure, we see that for the given value of $a$, with increasing $|h|$, the shadow size becomes larger and more distorted in comparison with the Kerr BH ($h=0$) in GR. We also see a horizontal shift in shadow along the $x$-axis, due to the frame dragging effect.

\section{Constraining parameters with the EHT observations}
\label{4-constrain}
In this section, we proceed to obtain some constraints on the plasma parameter using the EHT observations of the shadow of M87*. As we see from Fig.~\ref{shadow}, the shadow size depends on the plasma parameters and thus the observational data of M87* help us to impose bounds on them. As reported by the EHT collaboration \cite{EventHorizonTelescope:2019dse, EventHorizonTelescope:2019ths}, the angular diameter of the observed shadow of M87* is $\Delta\theta=(42 \pm 3)$ $\rm \mu as$, the mass and the distance from Earth to M87* are $M=(6.5\pm0.9)\times 10^{9}M_{\odot}$ and $D=(16.8\pm0.8)\rm Mpc$, respectively. Based on these data, one can estimates the diameter of the BH shadow, $d_{\rm sh}$, as \cite{Bambi:2019tjh}
\begin{equation}
d_{\rm sh}^{M87*}\equiv\frac{D\Delta\theta}{M}\approx11.0\pm1.5.\label{c1}
\end{equation}

\begin{figure}[H]
\centering
\includegraphics[width=3.10in]{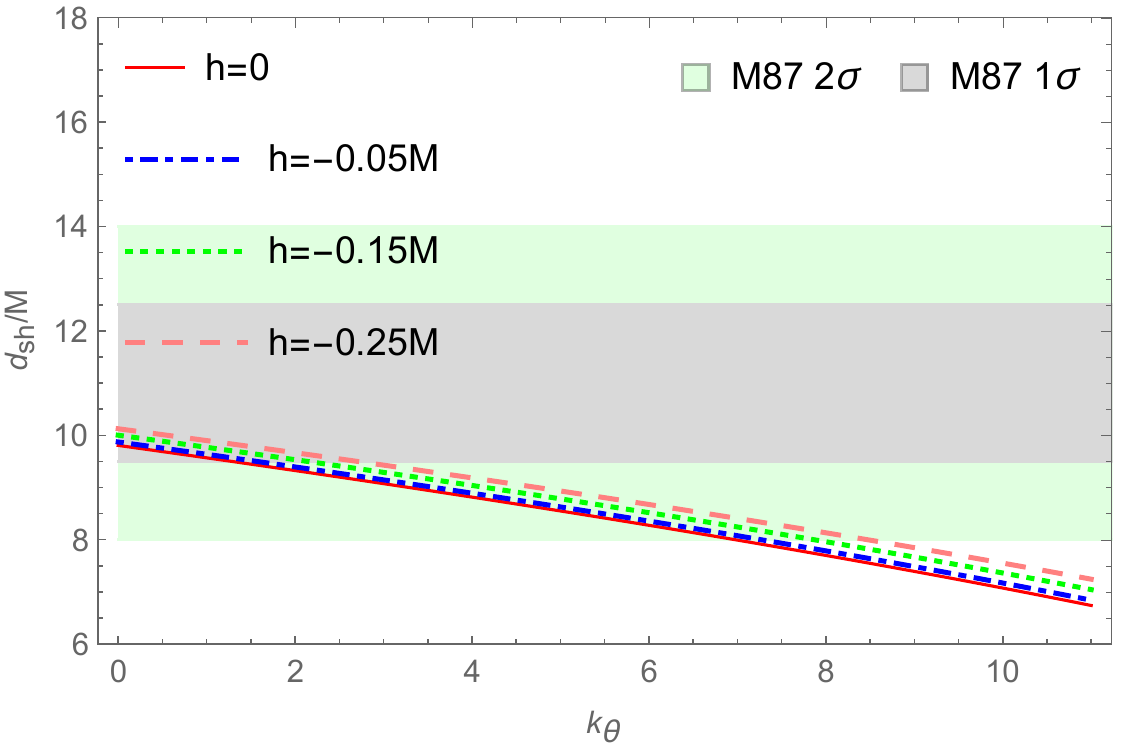}
\includegraphics[width=3.10in]{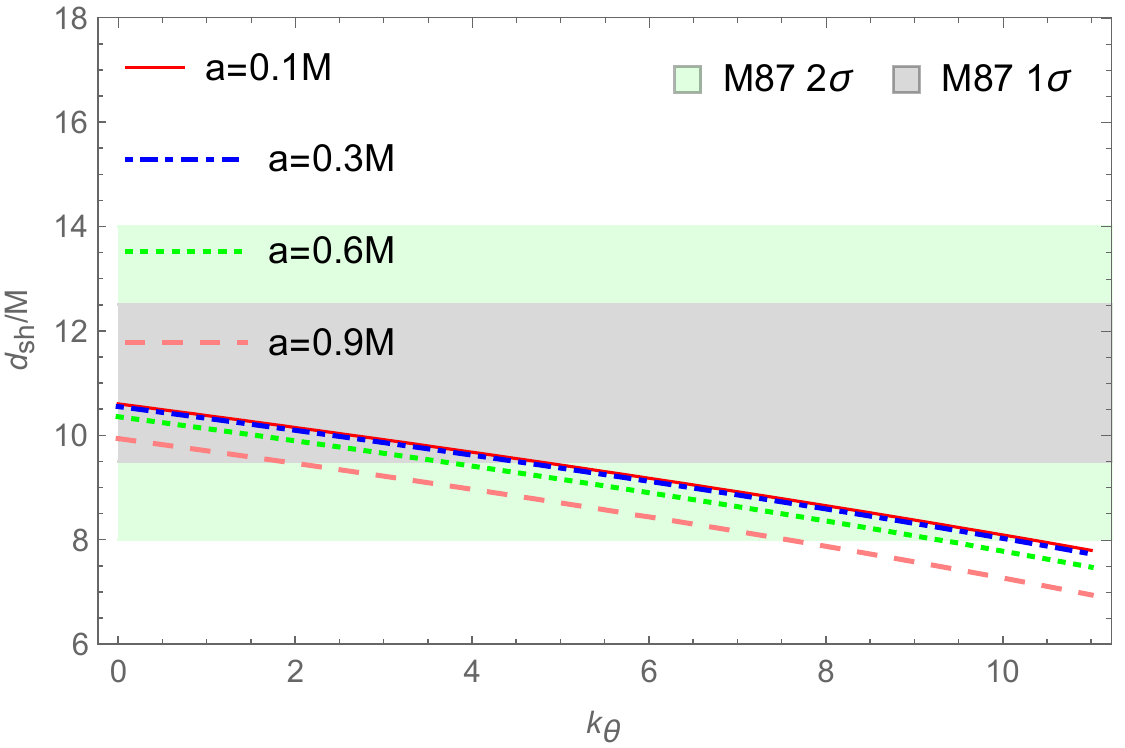}
\caption{\footnotesize The shadow diameter for the rotating hairy Horndeski BH as a function of longitudinal plasma parameter $k_{\theta}$, for fixed $a=0.9M$ and different values of hair parameter (left panel), and for fixed $h=-0.1M$ and different values of rotation parameter (right panel), with $\theta_{\rm o}=17^{\circ}$. In each panel the shaded area shows the observationally diameter of M87* shadow, within $1\sigma$ and $2\sigma$ confidence intervals.}
\label{c3}
\end{figure}

In order to match our theoretical results with these observational data, we need to calculate the shadow diameter $d_{\rm sh}$ of rotating Horndeski BHs. In our analysis, the inclination angle is taken to be $\theta_{\rm o}=17^{\circ}$, in agreement with EHT observations of M87* \cite{EventHorizonTelescope:2019dse}. In Fig.~\ref{c1} to Fig.~\ref{c3} we have displayed the behavior of the shadow diameter as a function of plasma parameters $k_{0}$, $k_{r}$ and $k_{\theta}$ for different values of $h$ and $a$ parameters. The gray region shows the $1\sigma$ confidence interval, while the green region represents the $2\sigma$ confidence interval. In each figure, the left panel shows the shadow diameter for different values of hair parameter $h$ with $a=0.9M$, while the right panel illustrates the behavior of shadow diameter for different values of spin parameter $a$ with $h=-0.1M$. As is clear, rotating Horndeski BH is able to describe the shadow of M87*, provided that plasma parameters $k_{0}$, $k_{r}$ and $k_{\theta}$ are constrained to specific values, dependent of values of $a$ and $h$ parameters.
Tab.~\ref{T1} shows the upper bound on the homogeneous plasma parameter $k_0$ within the $1\sigma$ and $2\sigma$ confidence levels. It can be seen that by increasing $|h|$, the upper bound on the $k_{0}$ decreases. However, the constrained values of plasma parameter $k_{0}$ increases as the spin parameter $a$ increases.

\begin{figure}[H]
\centering
\includegraphics[width=3.10in]{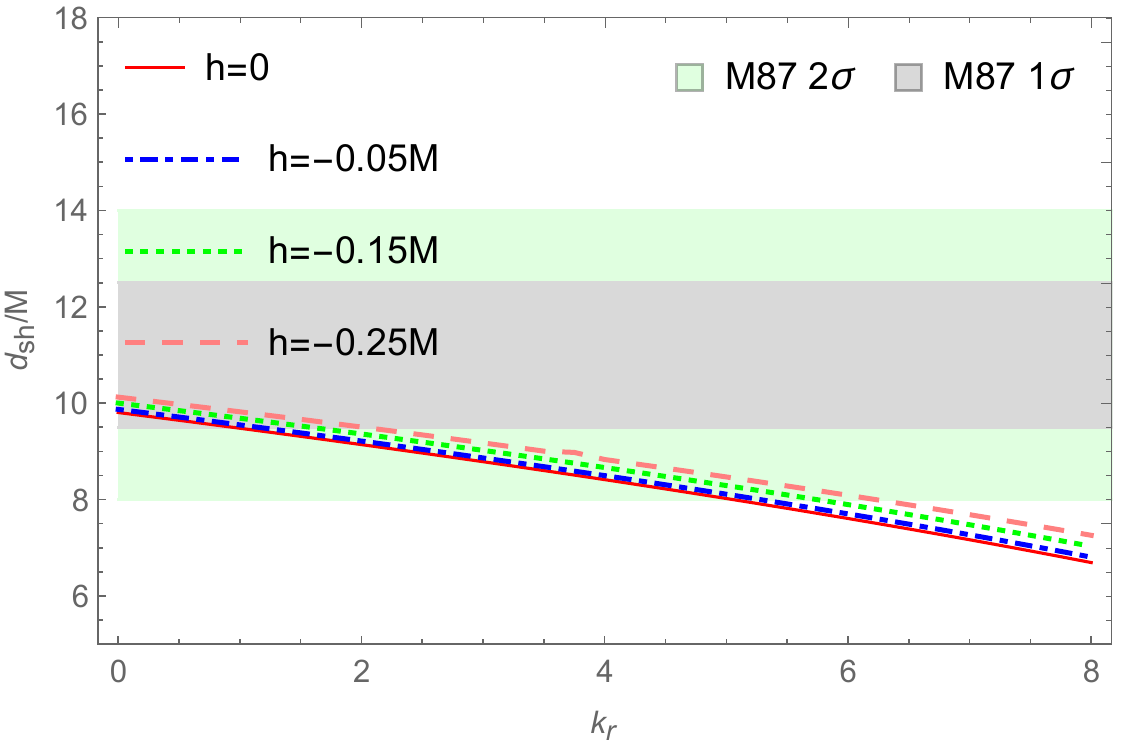}
\includegraphics[width=3.10in]{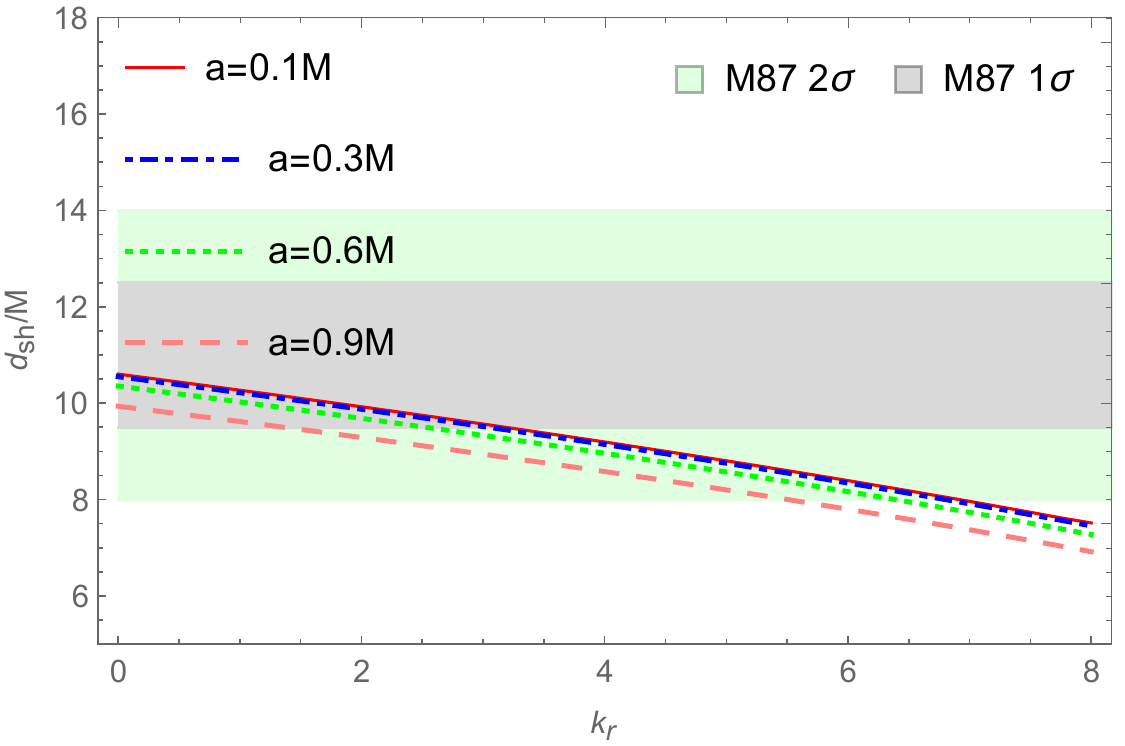}
\caption{\footnotesize The shadow diameter for the rotating hairy Horndeski BH as a function of radial plasma parameter $k_{r}$, for fixed $a=0.9M$ and different values of hair parameter (left panel), and for fixed $h=-0.1M$ and different values of rotation parameter (right panel), with $\theta_{\rm o}=17^{\circ}$. In each panel the shaded area shows the observationally diameter of M87* shadow, within $1\sigma$ and $2\sigma$ confidence intervals.}
\label{c2}
\end{figure}

We have also presented the results of constraints on the radial and latitudinal plasma parameters $k_{r}$ and $k_{\theta}$ in Tabs.~\ref{T2} and ~\ref{T3}, respectively. As it is clear from them as well as from the Fig.~\ref{c2} and Fig.~\ref{c3}, for a fixed value of $a$ with increasing $|h|$, the upper bound on the $k_{r}$ and $k_{\theta}$ increases too. Moreover, in contrast to the homogeneous plasma profile, the constrained values of plasma parameters $k_{r}$ and $k_{\theta}$ decrease as the spin parameter $a$ increases. Also, the comparison of results in Tab.~\ref{T2} and ~\ref{T3} shows that for the same values of $a$ and $h$ parameters, the bound on the radial plasma parameter $k_{r}$ is tighter than on the latitudinal plasma parameter $k_{\theta}$.

\begin{figure}[H]
\centering
\includegraphics[width=3.10in]{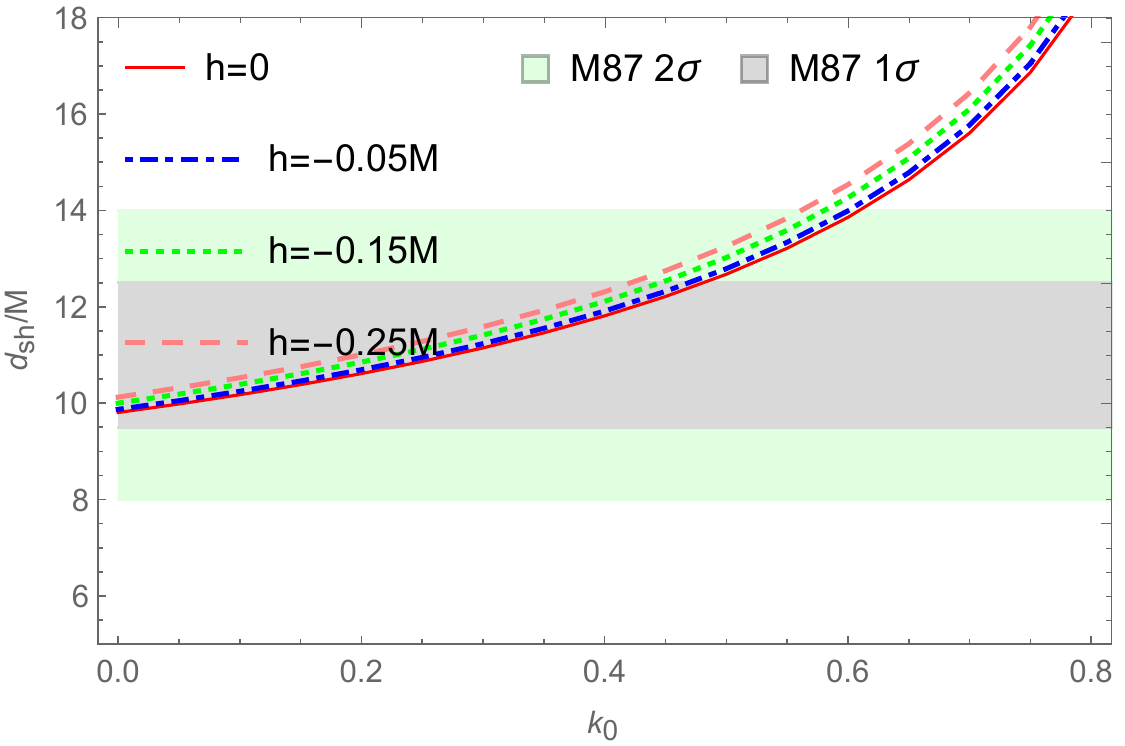}
\includegraphics[width=3.10in]{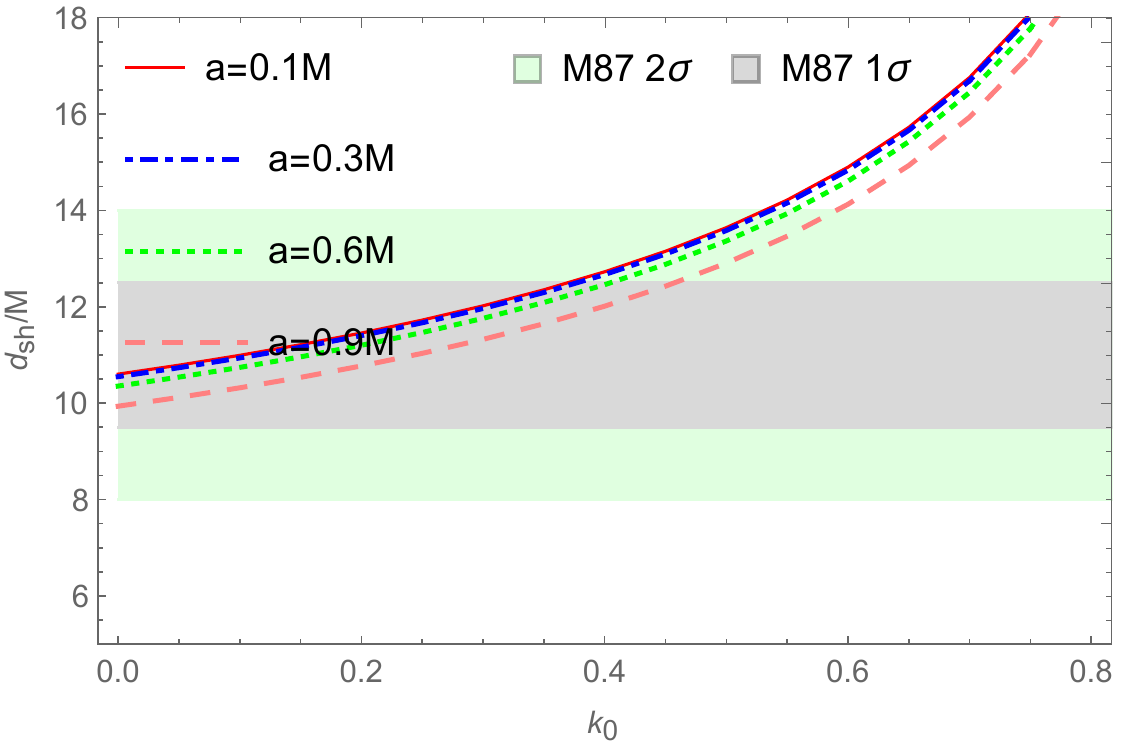}
\caption{\footnotesize The shadow diameter for the rotating hairy Horndeski BH as a function of homogeneous plasma parameter $k_{0}$, for fixed $a=0.9M$ and different values of hair parameter (left panel), and for fixed $h=-0.1M$ and different values of rotation parameter (right panel), with $\theta_{\rm o}=17^{\circ}$. In each panel the shaded area shows the observationally diameter of M87* shadow, within $1\sigma$ and $2\sigma$ confidence intervals.}
\label{c1}
\end{figure}

\begin{table}[H]
\centering
\caption{\footnotesize Constraints on plasma parameter $k_{0}$ from M87* data, for homogeneous plasma distribution with $\omega_p^2= k_0\omega_0^2$.}
\begin{tabular}{|c c c c||c c c c |c|}
\hline
$\shortstack{$h$ \\ $\quad\quad\quad$}$& $\shortstack{$a$ \\ $\quad\quad\quad$}$& \shortstack{upper bound $k_{0}$ \\ $1\sigma$}& \shortstack{upper bound $k_{0}$ \\ $2\sigma$}& $\shortstack{$h$ \\ $\quad\quad\quad$}$& $\shortstack{$a$ \\ $\quad\quad\quad$}$& \shortstack{upper bound $k_{0}$ \\ $1\sigma$}&  \shortstack{upper bound $k_{0}$ \\ $2\sigma$}\\ [0.5ex]
\hline
0       & $0.9M$& 0.481840& 0.609991& $-0.1M$& $0.1M$& 0.370356& 0.532056\\
$-0.05$M& $0.9M$& 0.469948& 0.600607& $-0.1M$& $0.3M$& 0.378045& 0.537211\\
$-0.15M$& $0.9M$& 0.446003& 0.581501& $-0.1M$& $0.6M$& 0.405133& 0.555373\\
$-0.25M$& $0.9M$& 0.422363& 0.562393& $-0.1M$& $0.9M$& 0.457987& 0.591082\\
\hline
\end{tabular}
\label{T1}
\end{table}

\begin{table}[H]
\centering
\caption{\footnotesize Constraints on plasma parameter $k_{r}$ from M87* data, for inhomogeneous plasma distribution with $\omega_p^2 = \frac{k_r\sqrt{r}}{r^2+a^2\cos^2\theta}\omega_0^2$.}
\begin{tabular}{|c c c c||c c c c |c|}
\hline
$\shortstack{$h$ \\ $\quad\quad\quad$}$& $\shortstack{$a$ \\ $\quad\quad\quad$}$& \shortstack{upper bound $k_{r}$ \\ $1\sigma$}& \shortstack{upper bound $k_{r}$ \\ $2\sigma$}& $\shortstack{$h$ \\ $\quad\quad\quad$}$& $\shortstack{$a$ \\ $\quad\quad\quad$}$& \shortstack{upper bound $k_{r}$ \\ $1\sigma$}&  \shortstack{upper bound $k_{r}$ \\ $2\sigma$}\\ [0.5ex]
\hline
0       & $0.9M$& 0.940303& 5.053828& $-0.1M$& $0.1M$& 3.182285& 6.912786\\
$-0.05$M& $0.9M$& 1.150365& 5.277357& $-0.1M$& $0.3M$& 3.044670& 6.802301\\
$-0.15M$& $0.9M$& 1.580306& 5.741222& $-0.1M$& $0.6M$& 2.528254& 6.388347\\
$-0.25M$& $0.9M$& 2.015712& 6.226327& $-0.1M$& $0.9M$& 1.364209& 5.506547\\
\hline
\end{tabular}
\label{T2}
\end{table}

\begin{table}[H]
\centering
\caption{\footnotesize Constraints on plasma parameter $k_{\theta}$ from M87* data, for inhomogeneous plasma distribution with $\omega_p^2 = \frac{k_{\theta}(1+2\sin^2\theta)}{r^2+a^2\cos^2\theta}\omega_0^2$.}
\begin{tabular}{|c c c c||c c c c |c|}
\hline
$\shortstack{$h$ \\ $\quad\quad\quad$}$& $\shortstack{$a$ \\ $\quad\quad\quad$}$& \shortstack{upper bound $k_{\theta}$ \\ $1\sigma$}& \shortstack{upper bound $k_{\theta}$ \\ $2\sigma$}& $\shortstack{$h$ \\ $\quad\quad\quad$}$& $\shortstack{$a$ \\ $\quad\quad\quad$}$& \shortstack{upper bound $k_{\theta}$ \\ $1\sigma$}&  \shortstack{upper bound $k_{\theta}$ \\ $2\sigma$}\\ [0.5ex]
\hline
0       & $0.9M$& 1.283236& 6.974963& $-0.1M$& $0.1M$& 4.722152& 10.314775\\
$-0.05$M& $0.9M$& 1.566179& 7.267375& $-0.1M$& $0.3M$& 4.488750& 10.084801\\
$-0.15M$& $0.9M$& 2.137535& 7.865668& $-0.1M$& $0.6M$& 3.637816& 9.257908\\
$-0.25M$& $0.9M$& 2.703517& 8.471147& $-0.1M$& $0.9M$& 1.851430& 7.564742\\
\hline
\end{tabular}
\label{T3}
\end{table}

\section{Conclusions}
\label{5-conclusion}
The astrophysical BHs in our universe are generally surrounded by plasma, and thus it is natural to explore the effects of plasma on the apparent shape of BHs. In the present work, we investigated the shadow of rotating hairy BHs surrounded by a plasma medium in the framework of Horndeski gravity. First, we studied the null geodesics in this environment by using the Hamilton-Jacobi equation. By considering both homogeneous and non-homogeneous plasma distributions, we discussed the Hamiltonian separability condition and obtained the celestial coordinates and the boundary of the Horndeski BH shadow for these plasma profiles. It is shown that the presence of a homogeneous plasma distribution causes the shadow size to increase in comparison to the vacuum case, while a non-homogeneous plasma profile causes it to decrease compared to that of the vacuum case, so that the BH shadow vanishes for higher plasma densities. Furthermore, investigating how the Horndeski parameter affects the BH shadow in the homogeneous plasma, showed that with increasing the absolute value of the hair parameter $|h|$, the shadow size becomes larger and more distorted in comparison with the Kerr BH ($h=0$) in GR. In fact, the result for this homogeneous plasma distribution is consistent with that in the vacuum case.
Finally, using observational data from M87*, we constrained the plasma parameters $k_{0}$, $k_{r}$ and $k_{\theta}$ within the $1\sigma$ and $2\sigma$ confidence levels based on the BH shadow diameter $d_{\rm sh}$, with the results summarized in Tabs.~\ref{T1} to ~\ref{T3}. Our findings showed that the rotating Horndeski BHs are able to describe the shadow size of M87*, provided that plasma parameters are constrained to specific values, dependent of values of $a$ and $h$ parameters.

\section*{Acknowledgments}
The authors would like to thank the Research Council of The University of Qom for financial support of project.

\end{document}